\documentclass[twocolumn,preprintnumbers, amssymb,amsmath,aps,floatfix,prl,nofootinbib,superscriptaddress,showpacs]{revtex4-1}\usepackage{mathrsfs}
\usepackage{amsfonts}
\usepackage{amsmath}
\usepackage{array}
\usepackage{verbatim}
\usepackage{epsfig}
\usepackage{graphicx}

\usepackage{xcolor}

\begin{document}
\title{Can We Observe Jet $P_T$-broadening in Heavy-Ion Collisions at the LHC?}

\author{Felix Ringer}
\affiliation{Physics Department, University of California, Berkeley, CA 94720, USA}
\affiliation{Nuclear Science Division, Lawrence Berkeley National
Laboratory, Berkeley, CA 94720, USA}

\author{Bo-Wen Xiao}
\affiliation{Key Laboratory of Quark and Lepton Physics (MOE) and Institute
of Particle Physics, Central China Normal University, Wuhan 430079, China}

\author{Feng Yuan}
\affiliation{Nuclear Science Division, Lawrence Berkeley National
Laboratory, Berkeley, CA 94720, USA}

\begin{abstract}
We investigate the effect of $P_T$-broadening on jet substructure observables in heavy-ion collisions at the LHC. As an example, we focus on the opening angle of the two branches that satisfy the soft drop grooming condition in a highly energetic jet. The medium modification of the angular distribution can provide important information on the jet transport properties of hot QCD matter. In addition, we take into account a change of the overall fraction of quark and gluon jets in heavy-ion collisions. We comment on the comparison to a recent measurement from the ALICE Collaboration.
\end{abstract}

\maketitle

{\it Introduction.}
Hard probes, in particular, highly energetic jets have become a powerful tool to study the medium properties of hot QCD matter created in heavy-ion collisions at RHIC and the LHC~\cite{Adcox:2001jp,Adler:2002xw,Aad:2010bu,Chatrchyan:2011sx}. The striking phenomenon of jet quenching has been addressed within different approaches to QCD parton energy loss in the literature~\cite{Gyulassy:1993hr,Baier:1996kr,Baier:1996sk, Baier:1998kq, Zakharov:1996fv,Gyulassy:1999zd, Wiedemann:2000za, Arnold:2002ja, Wang:2001ifa, CasalderreySolana:2010eh,Ovanesyan:2011xy, Casalderrey-Solana:2014bpa,Burke:2013yra,Zhang:2019toi}. For example, the approach by Baier-Dokshitzer-Mueller-Peigne-Schiff (BDMPS)~\cite{Baier:1996kr, Baier:1996sk, Baier:1998kq}, also predicts a significant transverse momentum broadening ($P_T$-broadening) of the jet due to multiple scatterings with the QCD medium. This $P_T$-broadening effect can be quantified by the so-called transport coefficient $\hat q$, where $\hat{q} L$ describes the typical transverse momentum squared that a parton acquires in the medium of length $L$. In the last few years, significant progress has been made in understanding $P_T$-broadening effects in hard scattering processes, including dijet, photon-jet, and hadron-jet correlations in heavy-ion collisions~\cite{Zapp:2012ak,Mueller:2016gko,Mueller:2016xoc,Chen:2016vem,Chen:2016cof,Chen:2018fqu,Tannenbaum:2017afg,Andrews:2018jcm,Adamczyk:2017yhe,Dominguez:2019ges}. The hadron-jet correlation measured by the STAR Collaboration has shown $P_T$-broadening effects at RHIC~\cite{Adamczyk:2017yhe}, whereas the dijet kinematics at the LHC are dominated by vacuum Sudakov effects and are thus not very sensitive to medium effects~\cite{Mueller:2016gko}. It is therefore important to search for other processes which are sensitive to $P_T$-broadening effects. The goal of this paper is to investigate this type of physics in the context of jet substructure measurements. 

A decade ago, jet substructure techniques were proposed as a method to study boosted topologies at the LHC focusing in particular on the discrimination of a boosted Higgs or electroweak gauge bosons from the QCD background~\cite{Butterworth:2008iy,Ellis:2009me,Krohn:2009th}. Since then significant further developments of jet substructure techniques have been achieved, see for example~\cite{Larkoski:2017jix}. Jet substructure observables also provide a unique opportunity to study various aspects of QCD dynamics in highly energetic jets. The substructure of jets has also generated great interest in the heavy-ion community where the experimental collaborations have for example reported results for several groomed observables~\cite{Sirunyan:2017bsd,Sirunyan:2018gct, Acharya:2019djg,Kauder:2017cvz}. Similarly on the theory side, jet substructure have received a growing attention~\cite{Mehtar-Tani:2016aco,Chien:2016led,Milhano:2017nzm,Chang:2017gkt,Li:2017wwc,Chien:2018dfn,Sirimanna:2019bgl,Caucal:2019uvr,Casalderrey-Solana:2019ubu,KunnawalkamElayavalli:2017hxo}. In this work, we study the possible impact of $P_T$-broadening effects due to multiple interactions of the particles inside the jet with the medium. As a first example, we focus on soft drop groomed jets~\cite{Larkoski:2014wba} and study the modification of the opening angle of the two branches satisfying the grooming condition which is also referred to as the groomed radius.  

When a highly energetic jet traverses the QCD medium, a $P_T$-broadening effect is generated due to multiple interactions. However, these effects will not affect the angular distribution of the two branches identified by the soft drop algorithm if the two branches interact coherently with the medium. Only if this interaction is incoherent, the medium effects will lead to a broadening of the internal angular distribution of the jet substructure. Therefore, the experimental measurement of this effect will provide crucial information on the interaction of jets with the medium and shed light on the underlying physics mechanisms at RHIC and LHC.

In order to compare to the experimental results of the jet substructure measurement, we also need to consider the overall jet energy loss. As demonstrated for example in recent studies~\cite{Qiu:2019sfj,He:2018gks,Brewer:2018dfs}, the energy loss of quark and gluon jets is different which leads to a relative change of the quark/gluon fractions in heavy-ion relative to proton-proton collisions. Since the angular distribution of the jet substructure observable considered here is different for quark and gluon jets, this change will automatically lead to a different shape of the substructure measurement even without jet $P_T$-broadening effects due to the medium.

In the following calculations, we take into account both effects, and comment on the comparison to a recent measurement by the ALICE Collaboration. We hope this will encourage further developments on both the experimental and theoretical side to build a comprehensive framework to study this type of physics. The remainder of this paper is organized as follows. We first briefly review the soft drop grooming algorithm and we show results for the opening angle of the two branches and include a nonperturbative model to account for hadronization effects. We then introduce $P_T$-broadening effects induced by multiple interactions of the jet and the medium. Finally, we summarize our results and present an outlook to further investigate the effects discussed in this work.

{\it Soft drop groomed jet substructure observables.} We start with an inclusive jet sample $pp\to{\rm jet}+X$ reconstructed using the anti-$k_T$ algorithm~\cite{Cacciari:2008gp} with jet radius $R$. The soft drop grooming procedure of~\cite{Dasgupta:2013ihk,Larkoski:2014wba} is designed to remove soft wide-angle radiation from the jet allowing for a more direct comparison of data and purely perturbative QCD calculations. Hadronization and underlying event contributions are significantly reduced. See for example the soft drop groomed jet mass distribution~\cite{Aaboud:2017qwh,Sirunyan:2018xdh,Frye:2016aiz,Marzani:2017mva,Kang:2018jwa}. The obtained jets are first reclustered using the Cambridge/Aachen (C/A) algorithm~\cite{Dokshitzer:1997in,Wobisch:1998wt} in order to obtain an angular ordered clustering tree. The C/A distance metric only depends on the distance of particles in the $\eta$-$\phi$ plane. The obtained jet is then declustered recursively where at each step one checks whether the two branches satisfy the soft drop grooming condition
\begin{equation}
    \frac{\min(p_{T1},p_{T2})}{p_{T1}+p_{T2}}>z_{\rm cut}\left(\frac{\Delta R_{12}}{R}\right)^\beta \, .
\end{equation}
Here $p_{T1,2}$ are the transverse momenta of the two branches at each declustering step and $\Delta R_{12}^2=\Delta\eta^2+\Delta\phi^2$ is their distance in the $\eta$-$\phi$ plane. There are two free parameters that define the soft drop declustering sequence which are the soft threshold $z_{\rm cut}$ and the angular exponent $\beta$. Note that in the limit $\beta\to\infty$ the groomer is removed and the ungroomed jet is recovered. Here we limit ourselves to $\beta=0$ which is currently the most common choice in heavy-ion data analyses. In principle, any jet substructure observable can now be measured on the remaining constituents of the groomed jet. In this work, we focus on the opening angle of the remaining branches or the so-called soft drop groomed radius $R_g$. In addition, the momentum sharing fraction of the two branches is denoted by $z_g$. Their definition is given by
\begin{eqnarray}
    \theta_g=\frac{\Delta R_{12}}{R}=\frac{R_g}{R} \, ,~~~~
    z_g=\frac{\min(p_{T1},p_{T2})}{p_{T1}+p_{T2}}\, .
\end{eqnarray}
Both observables have interesting features making them relevant for an exploration in the heavy-ion environment. Up to power corrections, the cumulative cross section $\Sigma(\theta_g)$ differential also in the jet transverse momentum $p_T$ and rapidity $\eta$ can be written as
\begin{equation}\label{eq:qgvac}
\frac{1}{\sigma_{\rm incl}}\frac{{\rm d}\Sigma(\theta_g)}{{\rm d}p_T \,{\rm d}\eta}=f_q\;\Sigma_q(\theta_g)+f_g\;\Sigma_g(\theta_g) \,,
\end{equation}
 where $\sigma_{\rm incl}$ denotes the inclusive jet cross section. Here, the $f_{i}$ denote quark/gluon fractions which are independent of the jet substructure observable $\theta_g$. In proton-proton collisions they can be calculated in terms of PDFs, hard-scattering and jet functions~\cite{Dasgupta:2014yra,Kaufmann:2015hma,Kang:2016mcy,Dai:2016hzf}. In heavy-ion collisions they get modified due to additional medium-induced out-of-jet radiations which can be obtained using a model calculation of parton energy loss or they can be determined phenomenologically from data~\cite{Qiu:2019sfj}. In general, a significant shift toward quark jets is expected in heavy-ion collisions which we estimate in our numerical results below. The $P_T$-broadening effects that we are going to discuss in the following only affect the functions $\Sigma_i(\theta_g)$ which resum in-jet radiations. The result for $\Sigma_i(\theta_g)$ at leading-logarithmic accuracy and fixed coupling can be written as~\cite{Larkoski:2014wba,Larkoski:2015lea,Larkoski:2017bvj,Tripathee:2017ybi}
\begin{eqnarray}
\Sigma_i(\theta_g)&=&\exp\Big[-C_i\frac{\alpha_s}{\pi}\int_{\theta_g}^1\frac{{\rm d}\theta}{\theta}\int_{z_{\rm cut}}^{1/2}{\rm d}z \, \overline{P}_i(z)\Big]  \, .
\end{eqnarray}
Here $C_i$ are the Casimir factors for quarks and gluons and $\overline{P}_i(z)$ denotes the respective collinear splitting functions summed over final states. We introduce the parameter $a_i$ as
\begin{equation}
    a_i=C_i\frac{\alpha_s}{\pi}\int_{z_{\rm cut}}^{1/2}{\rm d}z \, \overline{P}_i(z) \, ,
\end{equation}
and we can then rewrite the differential cross section in $\theta_g$ for quarks and gluons as
\begin{equation}
\frac{{\rm d}}{{\rm d}\theta_g}\Sigma_i(\theta_g) =\frac{a_i}{\theta_g} e^{a_i\ln\theta_g}=a_i\theta_g^{a_i-1}\, .
\end{equation}
The $\theta_g$ distribution was studied in the context of the CMS Open data analysis of~\cite{Tripathee:2017ybi}. Especially at small-$\theta_g$, a nonperturbative contribution is required. Since the following discussion of jet $P_T$-broadening effects in heavy-ion collisions depends on the physics in the entire kinematic range, we need to model nonperturbative effects also in proton-proton collisions.

A source of nonperturbative contributions are hadronization corrections which are particularly relevant at small $\theta_g$. We note that the underlying event contribution is expected to be very small due to the grooming procedure. To model hadronization effects, we include an additional transverse momentum kick perpendicular to the direction of each branch. We further assume that this effect acts independently on each of the branches with a constant transverse momentum squared (order of a few $\rm GeV^2$). We can then estimate the average $\langle \theta_\perp^2 \rangle$ angular deflection as
\begin{equation}\label{eq:q02}
    \langle\theta_\perp^2\rangle R^2=\frac{Q_0^2}{p_T^2z_g^2} \, ,
\end{equation}
where $p_T$ is the jet transverse momentum which is of the same order as $p_{T1}+p_{T2}$ of the two branches~\cite{Larkoski:2014wba}, $z_g$ is the momentum fraction of one of the branches, and $Q_0^2$ is the typical transverse momentum squared acquired by the branch due to the hadronization process. In the following, we assume a simple Gaussian distribution
\begin{equation}
    \frac{{\rm d}^2N}{{\rm d}^2\theta_\perp}=\frac{1}{\pi \Theta_s^2}e^{-\frac{\theta_\perp^2}{\Theta_s^2}} \, ,
\end{equation}
which is normalized to unity upon integration and where $\Theta_s^2$ is defined as
\begin{equation}\label{eq:q02p}
    \Theta_s^2=C_i\frac{Q_0^2}{p_T^2 R^2}\left(\frac{1}{z_g^2}+\frac{1}{(1-z_g)^2}\right) \, . 
\end{equation}
Here we introduced an average color factor to account for the flavor dependence of the nonperturbative contribution: $C_q=(C_F+C_A)/2C_F$ for a quark and $C_g=C_A/C_F$ for a gluon jet, respectively. The final $\theta_g$ distribution can be written as
\begin{eqnarray}
    \frac{{\rm d}\sigma}{{\rm d}\theta_g}&=&\int {\rm d}\theta_0 \, {\rm d}^2\theta_\perp \frac{{\rm d}\Sigma(\theta_0)}{{\rm d}\theta_0} 
    \frac{1}{\pi \Theta_s^2}e^{-\frac{\theta_\perp^2}{\Theta_s^2}} \delta(\theta_g-\ldots) \,\nonumber\\
   &\approx & \int {\rm d}^2\theta_\perp \frac{\theta_g}{\theta_0}\frac{{\rm d}\Sigma(\theta_0)}{{\rm d}\theta_0} 
    \frac{1}{\pi \Theta_s^2}e^{-\frac{\theta_\perp^2}{\Theta_s^2}}\, ,
\end{eqnarray}
where $\theta_g=\sqrt{\theta_0^2+\theta_\perp^2+2\theta_0\theta_\perp\cos\phi}$ and we left the dependence on $p_T,\,\eta$ and additional integrals over $z_g,\phi$ implicit. Note that we consider hadronization effects separately for the two branches since in general they can have different energies.

\begin{figure}
\includegraphics[width=8cm]{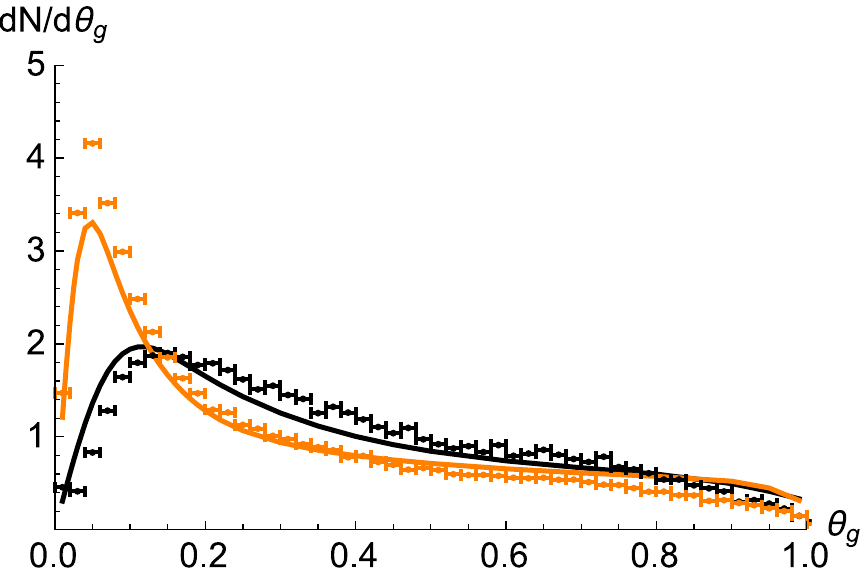}
\caption{The normalized cross section differential in the groomed radius $\theta_g=R_g/R$ in proton-proton collisions at 7~TeV including nonperturbative effects compared to results from the CMS open data analysis of~\cite{Tripathee:2017ybi} with $R=0.5$, $|\eta|<2.4$ and transverse momenta $p_T=85-115\; \rm GeV$ (black) and $p_T=200-250\;\rm GeV$ (orange).}
\label{CMS0}
\end{figure}

In Fig.~\ref{CMS0}, we show a comparison of the normalized $\theta_g$-differential distribution including nonperturbative effects in comparison to results from the CMS open data analysis of~\cite{Tripathee:2017ybi} at 7~TeV. Jets were identified with $R=0.5$, $|\eta|<2.4$ and in two $p_T$ intervals $p_T=85-115\;\rm GeV$ (black) and $p_T=200-250\;\rm GeV$ (orange). We use the parameter $Q_0^2=1.54$~GeV$^2$ for both $p_T$ intervals. With the nonperturbative contribution, the CMS Open data results are reasonably well described. However, we note that the results are not corrected for detector effects~\footnote{We also compared to the unfolded preliminary data in $pp$ collisions from ALICE at $\sqrt{s}=7\rm TeV$ at the LHC~\cite{Andrews:2018wgw} and STAR at $\sqrt{s}=200\rm GeV$ at RHIC~\cite{KunnawalkamElayavalli:2019wrv}, and found reasonably good agreement as well.}. Our results provide the baseline for the studies of the $\theta_g$ distribution in heavy-ion collisions discussed in the next section.

{\it Jet Substructure Observables in Heavy-Ion Collisions.} We assume that the heavy-ion cross section for the $\theta_g$ distribution can be cast in a similar form as the factorization structure at leading power in proton-proton collisions shown in  Eq.~(\ref{eq:qgvac}). As mentioned in the {\it Introduction}, we have to consider two effects: First, the quark/gluon fractions $f_{q,g}$ in Eq.~(\ref{eq:qgvac}) can change and, second, the internal structure for the quark and gluon jets encoded in the functions $\Sigma_i(\theta_g)$ can be different in heavy-ion collisions as well.

\begin{figure}
\includegraphics[width=8.5cm]{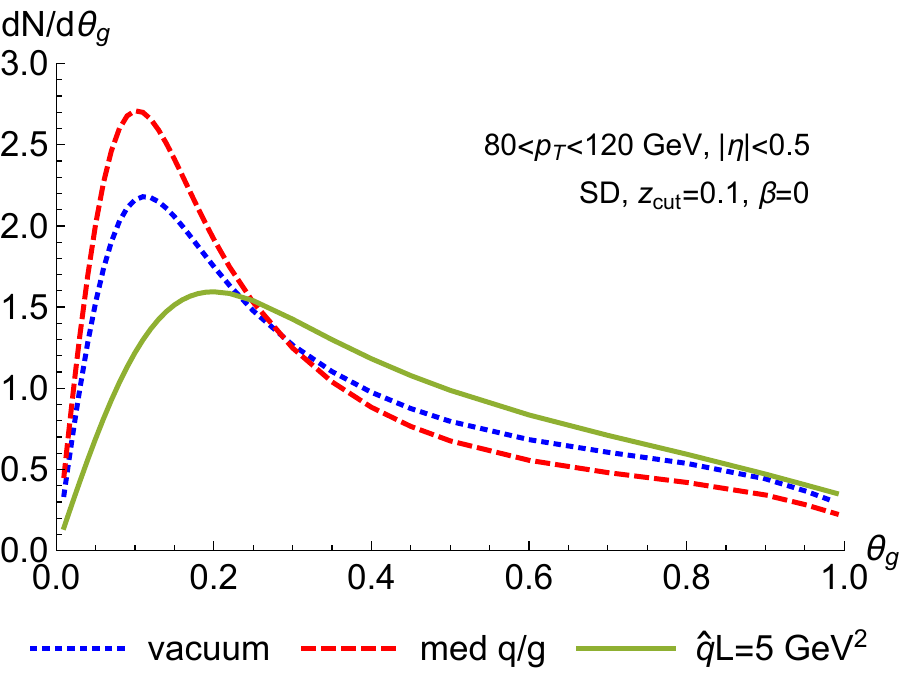}
\caption{The $\theta_g$ distribution for ALICE kinematics in the vacuum (dotted blue); with in-medium quark/gluon fractions (dashed red) or $P_T$-broadening effects with $\langle \hat qL\rangle=5$~GeV$^2$ (solid green). 
\label{fig:ALICE-qg_broadening}}
\end{figure}

We start with the modification of the overall quark/gluon fractions~\cite{Banfi:2006hf,Cal:2019hjc} due to interactions with the medium. For this, we follow a recent study based on QCD factorization for inclusive jet production in heavy-ion collisions~\cite{Qiu:2019sfj}. In this paper, the in-medium quark/gluon fractions were extracted within a global analysis of inclusive jet production data ${\rm PbPb}\to {\rm jet}+X$ at the LHC using the same factorization structure as in proton-proton collisions. A significant shift toward quark jets in the final state inclusive jet sample in heavy-ion collisions was observed. Qualitatively such a shift is expected as gluons lose more energy than quarks. Several jet substructure observables also suggest a large shift toward quark jets~\cite{Acharya:2018uvf,Aaboud:2018hpb,Sirunyan:2018qec,Spousta:2015fca}. The overall shift toward quark jets is illustrated in Fig.~\ref{fig:ALICE-qg_broadening}. The vacuum result is shown by the dotted blue curve for the ALICE kinematics of~\cite{Acharya:2019djg} at $\sqrt{s}=5.02$~GeV with jet kinematics $80<p_T<120$~GeV, $|\eta|<0.5$ and radius $R=0.4$ and the soft drop parameters $z_{\rm cut}=0.1$, $\beta=0$. Since quark jets peak at smaller values of $\theta_g$, we observe an overall narrowing of the distribution (dashed red) relative to the vacuum.

Next we consider $P_T$-broadening effects due to incoherent multiple scatterings of the two branches inside the medium, which generally leads to an opposite effect compared to the narrowing discussed above. First of all, these two branches passing the soft drop criterion are separated by a distance larger than the Debye length ($\sim 1\text{GeV}^{-1}$), which means that they can see different color charges in the medium. In addition, since the relative momentum between these two branches is much less than their total momentum, one can show that multiple scatterings of the two branches and medium can be approximately factorized into two independent scattering amplitudes (a similar example can be found in~\cite{Lappi:2012nh}). Following the BDMPS mechanism, this will lead to a $P_T$-broadening effect for each of the branches. Therefore, the angle or distance between the branches will get modified accordingly. Following the arguments of~\cite{Mueller:2016gko,Mueller:2016xoc}, the effects of multiple interactions with the medium do not affect the QCD resummation formalism in the vacuum as discussed in the previous section.  In the following, we parametrize the $P_T$-broadening effects in terms of the color charge of the parent jet: $(C_F+C_A)/2$ for a quark jet and $C_A$ for a gluon jet. We note that the $P_T$-broadening effects occur in the direction perpendicular to the large momentum component of the branches. This effect is thus similar to the hadronization effect in proton-proton collisions considered in the previous section. In heavy-ion collisions we thus replace $Q_0^2$ in Eqs.~(\ref{eq:q02}) and~(\ref{eq:q02p}) with 
\begin{equation}\label{eq:shift}
    Q_0^2+ f(z_{\rm cut},\beta)\,\langle \hat qL\rangle \, ,
\end{equation}
where we include the average transverse momentum broadening parametrized in terms of $\langle \hat qL\rangle$. In addition, we include a function $f(z_{\rm cut},\beta)$ that parametrizes the dependence on how aggressively the grooming algorithm removes soft radiation inside the jet. The functional dependence of $f(z_{\rm cut},\beta)$ could be studied more systematically by using the techniques developed in~\cite{Hoang:2019ceu}. To illustrate the effect of $P_T$-broadening on the $\theta_g$ distribution, we show the result with a typical $\langle \hat qL\rangle =5$~GeV$^2$~\cite{Burke:2013yra} in Fig.~\ref{fig:ALICE-qg_broadening} (solid green) without modifying the quark/gluon fractions and we choose $f(z_{\rm cut},\beta)=1$ for illustrative purposes. 
As expected, the $P_T$-broadening effects push the $\theta_g$ distribution toward higher values. This is a direct consequence of incoherent multiple scatterings of the two-branch system with the medium. Again, if the multiple interactions are coherent, it will only affect the total transverse momentum of the system and there will be no $P_T$-broadening effects visible in the $\theta_g$ distribution. 

If we consider the double differential cross section in $\theta_g$ and $z_g$, an additional test of $P_T$-broadening effects could be obtained. From Eqs.~(\ref{eq:q02p}) and~(\ref{eq:shift}) we see that $\Theta_s^2$ increases with decreasing $z_g$. As a result, we expect stronger $P_T$-broadening effects for smaller $z_g$ values.

{\it Comment on the Comparison to the ALICE measurement.} At the LHC, both the CMS and ALICE collaborations have reported measurements of soft drop groomed jet substructure observables~\cite{Sirunyan:2017bsd,Acharya:2019djg} focusing mostly on the $z_g$ distribution and its modification in the medium as well as the groomed mass~\cite{Sirunyan:2018gct}. 

However, ALICE reported the $z_g$ results for different $\Delta R_{12}$ cuts and provided corresponding ``tagged rates'' of the jets that satisfy the soft drop criterion in heavy-ion collisions using different $\Delta R_{12}$ intervals. In this sense the ALICE measurement already contains information on the $\theta_g$ distribution. The data from~\cite{Acharya:2019djg} can be classified into three $\theta_g$ bins: $\theta_g<0.25$ ($\Delta R_{12}<0.1$), $\theta_g>0.25$ ($\Delta R_{12}>0.1$) and $\theta_g>0.5$ ($\Delta R_{12}>0.2$). It is interesting to note that the distribution in heavy-ion collisions has significant support at large values of $\theta_g$. For example, roughly half of the events are in the $\theta_g>0.5$ bin. Compared to Fig.~\ref{fig:ALICE-qg_broadening}, this may indicate $P_T$-broadening effects. In addition, for different $z_g$-bins, we observe that the average $\theta_g$ decreases with $z_g$, which may be another indication of $P_T$-broadening effects. However, as pointed out in the ALICE paper~\cite{Acharya:2019djg}, we note that the published heavy-ion data are not unfolded. The data are also reasonably well described by the smeared results from the Jewel event generator~\cite{Zapp:2012ak}, where multiple interactions with the medium are included in the simulation. 

In addition, the ALICE paper shows a comparison of their results to an embedded Pythia simulation. Given the current accuracy of the data this comparison suggests that $P_T$-broadening effects are relatively small for this observable and the dominant effect is the relative change of the quark/gluon fractions in the medium.

Together with recent studies in Refs.~\cite{Caucal:2019uvr,Casalderrey-Solana:2019ubu}, the above discussions demonstrate that jet substructure measurements in heavy-ion collision with and without grooming have great potential to reveal the underlying QCD dynamics of the interaction of a jet with the hot and dense medium created at the LHC. We expect that the experimental uncertainties can be reduced in the future and a we hope that a dedicated $\theta_g$ analysis can be performed in order to reach more definitive conclusions. 

{\it Summary and Outlook.} We have presented a calculation of the opening angle $\theta_g$ of the two branches in a jet that pass the soft drop criterion. We find reasonable agreement with the CMS-open data results of the $\theta_g$ distribution after including a model of nonperturbative effects. We investigated the impact of an overall change of the quark/gluon fractions in heavy-ion collisions as well as $P_T$-broadening effects. The two effects lead to a narrowing or broadening of the $\theta_g$ distribution, respectively.  

With more precise data becoming available in the future it will be possible to reliably pin down the different medium effects. On one hand, the extraction of quark/gluon fractions~\cite{Qiu:2019sfj} needs to be improved. A complementary approach to obtain in-medium quark/gluon fractions was developed in~\cite{Brewer:2018dfs}. On the other hand, we expect significant new insights if the experimental collaborations can provide the nuclear modification factor using different grooming parameters which will give a better access to the soft dynamics of jets in heavy-ion collisions and potentially reveal $P_T$-broadening effects more clearly using jet substructure observables.

Furthermore, it will be interesting if an additional observable can be measured such as~\cite{Larkoski:2017cqq} which allows for the classification of jets into a one- or two-prong substructure. We hope that our studies encourage further developments in this direction which may eventually provide an effective technique to measure $\langle \hat q L\rangle$ of the hot and dense QCD matter created in heavy-ion collisions.

{\it Acknowledgement.} We wish to dedicate this work to the 80th birthday of Prof. Alfred Mueller, who has inspired and guided us in many ways. We thank Aashish Tripathee and Wei Xue for help with the CMS-open data analysis of Ref.~\cite{Tripathee:2017ybi}. We thank Leticia Cunqueiro, Peter Jacobs, Mateusz Ploskon and James Mulligan for discussions and comments concerning the ALICE measurements. We also thank Raghav Elayavalli, Yen-Jie Lee, Ben Nachman and Nobuo Sato for helpful discussions. The material of this paper is based upon work partially supported by the LDRD program of Lawrence Berkeley National Laboratory, the U.S. Department of Energy, Office of Science, Office of Nuclear Physics, under contract number DE-AC02-05CH11231. F.R.~is supported by the NSF under Grant No. ACI-1550228 within the JETSCAPE Collaboration.

\bibliographystyle{h-physrev}
\bibliography{bibliography}

\end{document}